# Coherent radiative decay of molecular rotations - - a comparative study of terahertz-oriented versus optically aligned molecular ensembles

Ran Damari, Dina Kardash, and Sharly Fleischer

Raymond and Beverly Sackler Faculty of Exact Sciences, School of Chemistry, Tel Aviv University 6997801, Israel.
Tel-Aviv University center for Light-Matter-Interaction, Tel Aviv 6997801, Israel
Email: sharlyf@post.tau.ac.il

Abstract:

The decay of field-free rotational dynamics is experimentally studied in two complementary methods: laser-induced molecular alignment and terahertz-field-induced molecular orientation. Comparison between the decay rates of different molecular species at various gas pressures reveals that oriented molecular ensembles decay faster than aligned ensembles. The discrepancy in decay rates is attributed to the coherent radiation emitted by the transiently oriented ensembles and is absent from aligned molecules. The experimental results reveal the dramatic contribution of coherent radiative emission to the observed decay of rotational dynamics and underline a general phenomenon expected whenever field-free coherent dipole oscillations are induced.

-----------------------------------------------------------------------------------------------------------

Gas samples are composed of individual molecules that are oriented with equal probability in all possible directions. Such molecular ensembles are termed 'isotropic' and lack any association between the lab- and molecular frames, thus entailing spectroscopic signals that are inherently averaged over all molecular-frame directions. Vast efforts are made toward lifting this inherent angular isotropy gas phase molecules by utilizing strong laser pulses that impart a short duration torque and rotate the molecules to align along a chosen lab-frame direction [1,2,3]. It is highly desired to spectroscopically interrogate the aligned/oriented sample under field-free conditions, namely long after the interaction with the aligning or orienting fields is over, to avoid any distortions or unwanted molecular responses that may be induced by the excitation fields. This may be achieved thanks to the quantization of angular momentum and correspondingly the rotational transition energies that manifest as periodic recurrences of aligned and/or oriented angular distributions. In linear molecules the periodicity at which the rotational wavepacket 'revives' is given by the inverse of the lowest transition frequency of the rotor $T_{rev}=(2B)^{-1}$, where B is the rotational constant in [Hz] [4,5].

Laser-induced alignment of linear molecules has been vastly explored since ultrashort laser pulses with milli-joule energies became tabletop-available by chirped pulse amplification technologies. Molecular alignment corresponds to an angular distribution of the molecular axes that is maximized along a specific lab-frame axis, while retaining the inversion symmetry of the medium at all times (dipoles point 'up' and 'down' with equal probability). Recently, significant efforts are made toward lifting the inversion symmetry of the gas medium to obtain molecular orientation - where the molecular dipole vectors are preferentially pointing toward a specific lab-frame direction. Both alignment and orientation are initiated via short duration molecule-field interaction that induces coherent rotational dynamics with periodic recurrences governed by the rotational spectrum of the molecules, the coherences induced via the interaction with the exciting field and the interactions between the sample molecules. The latter lead to exponential decay of the observed rotational dynamics, primarily governed by the rate of inter-molecular collisions. While alignment can be induced by a short laser pulse (typically 800nm), orientation requires either a two-color field ($\omega+2\omega$) [6,7], a mixed dc+optical excitation [8,9,10] or via resonant interaction with a single- or a few-cycle terahertz field [11,12,13,14].

In this work we measure and compare the decay rates of THz-induced orientation and optical-induced alignment and unveil a coherent radiative decay mechanism that, to the best of our knowledge, has been discarded from previous works on molecular orientation. As will be shown later, this coherent radiative decay imparts significant ramifications to the observed field-free decay of orientation dynamics. The results agree with recent observations of abrupt rotational energy loss in THz-induced field-free molecular orientation [15], however here, from the viewpoint of rotational coherences.

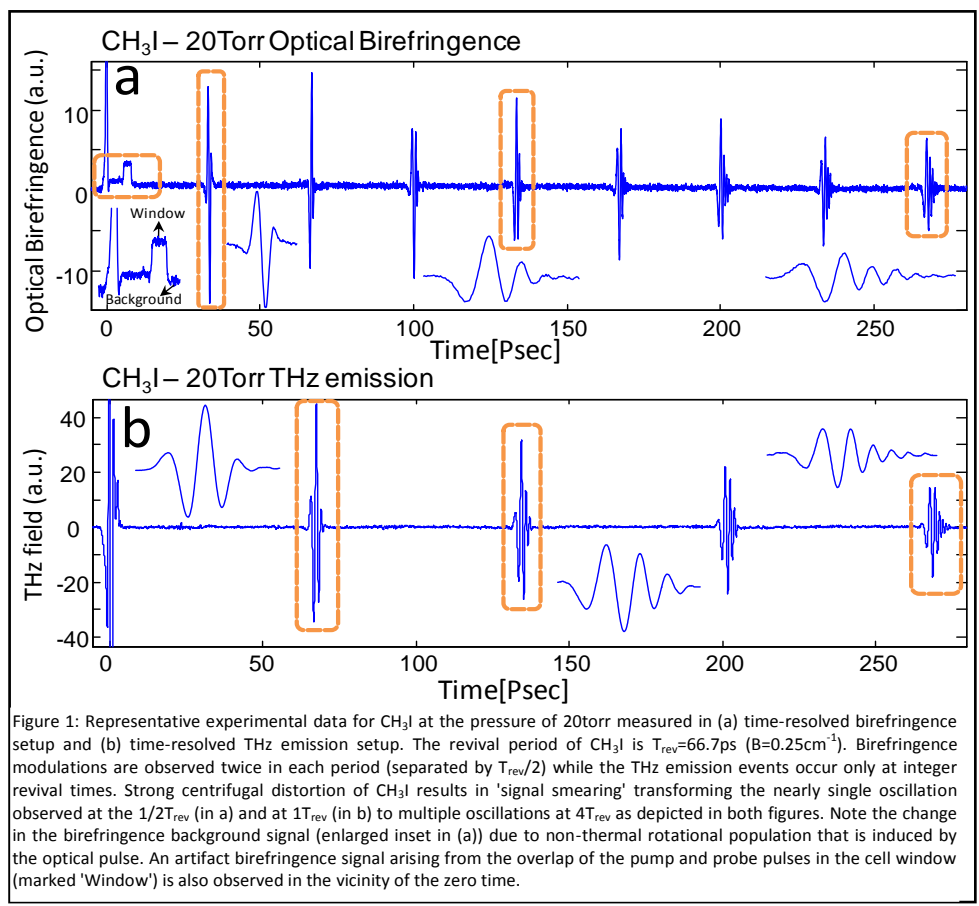

Figure 1: Representative experimental data for $CH_3I$ at the pressure of 20torr measured in (a) time-resolved birefringence setup and (b) time-resolved THz emission setup. The revival period of $CH_3I$ is $T_{rev}$=66.7ps (B=0.25cm$^{-1}$). Birefringence modulations are observed twice in each period (separated by $T_{rev}/2$) while the THz emission events occur only at integer revival times. Strong centrifugal distortion of $CH_3I$ results in 'signal smearing' transforming the nearly single oscillation observed at the $1/2T_{rev}$ (in a) and at $1T_{rev}$ (in b) to multiple oscillations at $4T_{rev}$ as depicted in both figures. Note the change in the birefringence background signal (enlarged inset in (a)) due to non-thermal rotational population that is induced by the optical pulse. An artifact birefringence signal arising from the overlap of the pump and probe pulses in the cell window (marked 'Window') is also observed in the vicinity of the zero time.

*Experimental -*

The rotational dynamics of three polar gases: N$_2$O (µ=0.166D), OCS(µ=0.7D) and CH$_3$I (µ=1.62D) were measured at a range of gas pressures in two different experimental setups: time-resolved optical birefringence and time-resolved THz emission setup, used previously in [16]. All of the measurements were performed in a 10cm static gas cell at the ambient temperature of the lab (295K). All gas samples were scanned for the entire time-span of the 250mm long computer-controlled delay stage (~1500ps usable delay range) and each data set was averaged over three full scans. Representative data of 20 Torr CH$_3$I as measured in the two experimental setups is shown in Fig.1, spanning the first four revival periods. Note that the THz-field emission signals shown in Fig.1b were obtained by deconvolution of the raw experimental data using the transfer function of the THz setup (the latter was experimentally measured with an evacuated gas cell). The deconvolution scheme is crucial for obtaining a clean molecular signal for further analysis and quantification by discarding various experimental artifact signals such as reflections in the gas-cell windows and in the electro-optic GaP/ZnTe detection crystal. For detailed description of the experimental setups and the deconvolution scheme see [17].

*Metrics for quantifying signal decay -*

Linear molecules can be modeled as quantum mechanical rigid rotors, with eigenfunctions given by the spherical harmonic functions $\left|Y_J^m\right\rangle$ and eigenenergies $E_{J,m} = BJ(J+1)$, with $B$ the rotational coefficient. The transition spectrum of the rigid rotor $\Delta E_{J \to J+1} = 2B(J+1)$ forms a purely harmonic series and manifest perfectly periodic recurrences of the excited rotational wavepacket and correspondingly identical signals at each revival event. However, due to the finite rigidity of real molecules, the rotational energies are slightly shifted due to the centrifugal force acting to stretch the molecule with increasing rotational energies. A correction term to the level energy, known as 'centrifugal distortion' is given by $DJ^2(J+1)^2$, with typical $B/D$ in the range of $10^5 - 10^7$. Therefore the corrected energy levels of the molecular rotor are given by $E_{J,m} = BJ(J+1) - DJ^2(J+1)^2$, resulting in slight incommensurability of the transition frequencies $\Delta E_{J \to J+1} = 2B(J+1) - 4D(J+1)^3$. The effects of centrifugal distortion are observed in figs. 1a and 1b. As time progresses the signals become longer in duration, with increasing number of oscillations and decreasing peak amplitudes (see enlarged insets of the alignment signals at 1/2T$_{rev}$, 2T$_{rev}$, and 4T$_{rev}$ and orientation signals at 1T$_{rev}$, 4T$_{rev}$). Since the peak amplitude of the revival signal changes with time, one must devise a metric, different from the peak signal amplitude, to quantify the signal strength - a metric that is insensitive to the changes in signal shape and amplitude imparted by the centrifugal distortion. Before we describe the metric schemes, we note that unlike $N_2O$ and OCS that are linear molecules, $CH_3I$ is a prolate symmetric-top with a very small moment of inertia about the C-I axis (3.26 amu·$A^2$) compared to the two other axes (67.4 amu·$A^2$). The ΔK=0 selection rule for the optical/THz excitation results in a qualitatively similar rotational dynamics to that of linear molecules [18,19,20]. Thus, for the sake of simplicity and uniformity of the theoretical treatment, $CH_3I$ is treated as a linear rotor throughout this work.

Two kinds of metrics/analysis schemes were examined for this purpose:

*metric 1* - Time domain integration over each revival signal separately.
*metric 2* - Integration over the frequency amplitude of each revival signal separately.
The integrated signals are fitted to a single exponent to extract the decay rate of the measured data. All of the data sets reported in this work were analyzed using both these metrics with excellent agreement between them as demonstrated in Fig. 2. Note that metric1 was used before by Owschimikow et.al. [21] where integration over the homodyne detected signals was fitted to exponential decay.

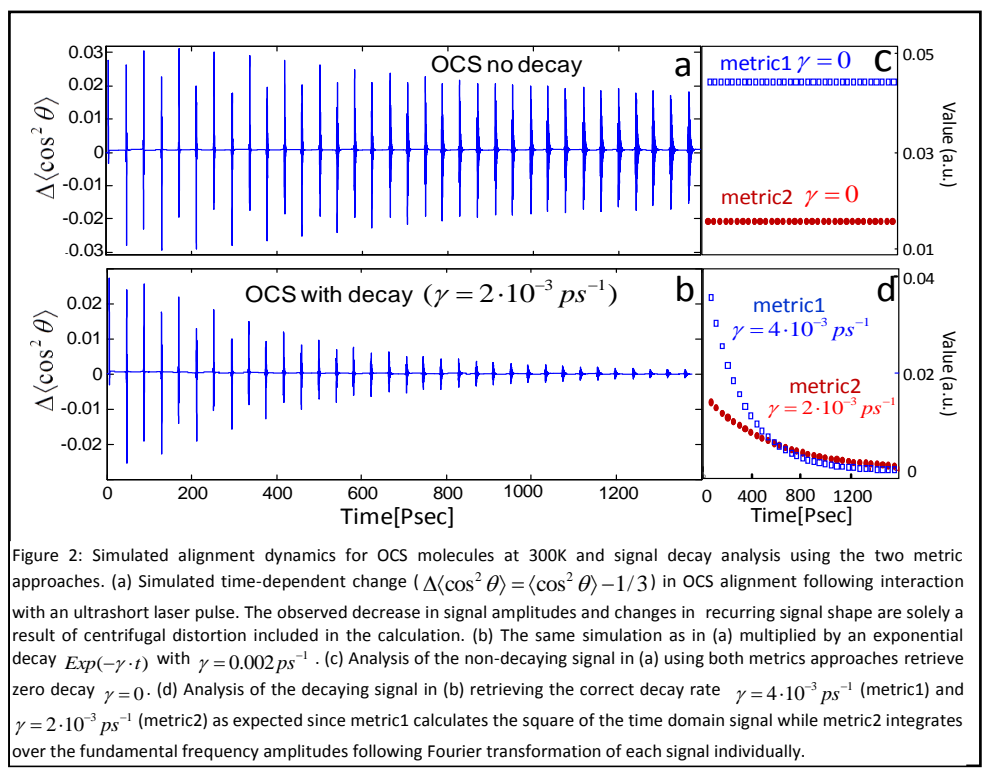

Figure 2: Simulated alignment dynamics for OCS molecules at 300K and signal decay analysis using the two metric approaches. (a) Simulated time-dependent change ($\Delta\langle\cos^2\theta\rangle = \langle\cos^2\theta\rangle - 1/3$) in OCS alignment following interaction with an ultrashort laser pulse. The observed decrease in signal amplitudes and changes in recurring signal shape are solely a result of centrifugal distortion included in the calculation. (b) The same simulation as in (a) multiplied by an exponential decay $Exp(-\gamma \cdot t)$ with $\gamma = 0.002 ps^{-1}$. (c) Analysis of the non-decaying signal in (a) using both metrics approaches retrieve zero decay $\gamma = 0$. (d) Analysis of the decaying signal in (b) retrieving the correct decay rate $\gamma = 4 \cdot 10^{-3} ps^{-1}$ (metric1) and $\gamma = 2 \cdot 10^{-3} ps^{-1}$ (metric2) as expected since metric1 calculates the square of the time domain signal while metric2 integrates over the fundamental frequency amplitudes following Fourier transformation of each signal individually.

Figure 2a shows a computer simulated decay- and decoherence- free dynamics of optical-induced alignment in carbonyl-sulfide (OCS) molecules ($B=0.203 cm^{-1}$, $D=4.33 \cdot 10^{-8} cm^{-1}$) extending up to 1.4ns. Including the centrifugal distortion term results in the observed decrease in the amplitude of the signals and the appearance of increasing number of oscillations. Fig.2c depicts the results of our analysis based on metric1 (blue squares) and metric2 (red circles) as a function of time. Fig.2b includes an exponential decay with a decay rate of $\gamma = 2 \cdot 10^{-3} ps^{-1}$, in addition to the centrifugal distortion, and is exactly retrieved by both metrics (with metric1 decaying twice faster as fast as metric2 due to the integration over the signal squared).

Armed with both metrics we now turn to examine and compare the field-free decay of oriented vs. aligned molecules. We performed time-resolved measurements of optical-induced alignment and THz-induced orientation of three molecular gases ($N_2O$, OCS, $CH_3I$). Each sample was measured at a range of pressures in a static gas cell at ambient temperature. For each sample we recorded a series of alignment and orientation signals like those shown in Fig.1, but for the entire time span provided by the computer controlled delay stage. To the best of our knowledge, previous works on optical-induced alignment attributed the decay of the signal to collisions among the gas sample constituents leading to thermalization of excited populations (J,M changing 'in-elastic' collisions) and decoherence induced via both 'elastic' and 'in-elastic' collisions with a buffer gas or among the sample constituents [21,22,23]. The collision-induced decay of coherent molecular rotations results in a single exponential decay that can be extracted by fitting the numerical and experimental results [11,19,24]. The decay of field-free orientation, however, has not been considered before and is at the focus of this work. We hypothesize that field-free oriented ensembles must decay faster than field-free aligned ensembles due to the emission of coherent radiation emerging upon molecular orientation. This radiation emission is absent from optical-induced alignment, hence the prospected difference in their decay rates. Our research strategy relies on the assumption that the collision-induced decay of the gas is indifferent to the type of field-excitation used to induce the rotational dynamics, i.e. whether the rotational dynamics is induced by a THz pulse interacting via the permanent dipole, $\hat{V}_{THz} = -\mu E \cos\theta$, or by an optical pulse interacting via the molecular polarizability $\hat{V}_{optical} = -\frac{1}{4}\Delta\alpha E^2(t)\cos^2\theta$ should not affect the rate of collisions and the corresponding decay of the signal. In both cases, once the excitation pulse is over (at t=100fs for optical pulse and t=2ps for the single-cycle THz field), the dynamics is governed by the field-free Hamiltonian and the collisions experienced by the gas molecules. Clearly this assumption may not hold for rotational excitations much stronger than ours, where the initial thermal distribution is shifted to highly excited states as in molecular super-rotors [25,26].

We performed measurements of the orientation and alignment dynamics and quantified the rotational decay rates of the two signals. The decay rates $\gamma_{Orient}$ and $\gamma_{Align}$ were extracted by fitting the experimental data, analyzed by the two metric approaches, to an exponential decay function $S(t) = S_0 \exp[-\gamma \cdot t]$. Within our working assumption mentioned above, $\gamma_{Align}$ corresponds to the pure collisional-decay rate ($\gamma_{Collision} \equiv \gamma_{Align}$), and $\gamma_{Orient}$ consists of two contributions: that of pure collisional-decay and the contribution of the radiative emission to the decay and decoherence, i.e. that $\gamma_{Orient} = \gamma_{Collisions} + \gamma_{Radiation}$. Thus, the optical-induced alignment measurement serves as reference for the collision-induced decay rate $\gamma_{Collisions}$ and enable the extraction of the radiative decay rate $\gamma_{Radiation}$.

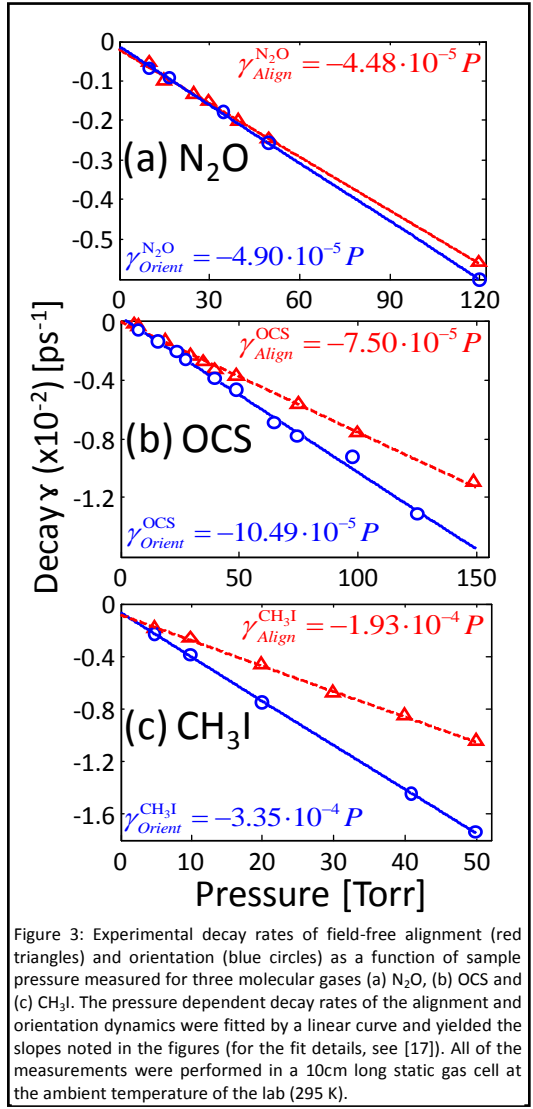

Figure 3: Experimental decay rates of field-free alignment (red triangles) and orientation (blue circles) as a function of sample pressure measured for three molecular gases (a) N$_2$O, (b) OCS and (c) CH$_3$I. The pressure dependent decay rates of the alignment and orientation dynamics were fitted by a linear curve and yielded the slopes noted in the figures (for the fit details, see [17]). All of the measurements were performed in a 10cm long static gas cell at the ambient temperature of the lab (295 K).

Figure 3 depicts the experimentally measured decay rates, $\gamma_{Align}$ (red triangles) and $\gamma_{Orient}$ (blue circles) as a function of gas pressure. Each data point was extracted from a long time-resolved scan (up to ~1.5ns past excitation time or in the range of decent S/N ratio) of the alignment (red) or orientation (blue) signal at the specified pressure. The decay rates are found to increase linearly with the gas density (linear with pressure), and are summarized in Table 1.

The dramatic contribution of radiative decay is evident from Fig.3 and the results summarized in Table 1, with $\gamma_{Align} < \gamma_{Orient}$ found for the three gasses. While for N$_2$O the decay slopes, $\gamma_{Align}$ and $\gamma_{Orient}$, are hardly deciphered, the difference is much more pronounced in OCS and is dramatically larger in CH$_3$I. Comparing the results from the three gasses, the contribution of $\gamma_{Radiation}$ clearly increases with the dipole moment µ.

| Gas | $\gamma_{Align}[(ps \cdot torr)^{-1}]$ | $\gamma_{Orient}[(ps \cdot torr)^{-1}]$ | $\gamma_{Radiation}[(ps \cdot torr)^{-1}]$ $= \gamma_{Orient} - \gamma_{Align}$ | $\gamma_{Radiation}[(T_{rev} \cdot torr)^{-1}]$ |
|---|---|---|---|---|
| N$_2$O (µ=0.16D) T$_{rev}$=39.8ps | 2.24e-5 | 2.45e-5 | 2.1e-6 | 8.36e-5 |
| OCS (µ=0.7D) T$_{rev}$=82ps | 3.75e-5 | 5.25e-5 | 1.5e-5 | 1.23e-3 |
| CH$_3$I (µ=1.62D) T$_{rev}$=67ps | 9.65e-5 | 16.75e-5 | 7.1e-5 | 4.76e-3 |

Table 1: Decay rates extracted from the fits in Fig.3. $\gamma_{Radiation}$ is extracted from the difference $\gamma_{Orient} - \gamma_{Align}$. Note that the decay rates are taken from the fit in Fig.3, only divided by 2 to account for the integration over the signal squared.

Theoretical model

In what follows, we suggest a theoretical model that includes a radiation emission term in the Hamiltonian. The typical Hamiltonian that describes the dipole-induced rotational dynamics relies on the quantum-mechanical rigid rotor model and is given by $\hat{H} = \hat{L}^2/2I + \hat{H}_{int}$, with $H_{int}(t) = -\mu \cdot E(t) \cdot \cos(\theta)$ is the dipole interaction term, *E(t)* is the time-dependent electric field, $\hat{L}$ the angular momentum operator, $I$ the moment of inertia, and $\mu$, the dipole moment of the molecule [27,28,29,30,31,32,20]. Following the short-time interaction with the field E(t), the dynamics is governed solely by the field-free term $\hat{L}^2/2I$, and the dynamics can, in principle persist forever under decay- and decoherence-free conditions. Polar molecules that are rotationally excited via field-dipole interaction, orient with each revival period, forming a 'phased array' of oriented molecules that emit coherent radiation (known as free-induction-decay signal, FID) in the forward direction (the direction of the THz excitation pulse). This FID emission serves to characterize the orientation dynamics of the molecular ensemble, however its ramifications on the molecular ensemble are typically discarded from the Hamiltonian. In order to account for the energy loss via emitted radiation we included an additional term in the Hamiltonian: $H_R(t) = \mu \cdot dP(t)/dt = \mu \cdot \frac{d(\mu\langle\langle\cos\theta\rangle\rangle)(t)}{dt}$, where $P(t)$ is the transient dipole $\mu\langle\langle\cos\theta\rangle\rangle(t)$ and its derivative with respect to time provides the radiated field [11,19]. At each time step of the simulation we calculate the ensemble-averaged orientation $\langle\langle\cos\theta\rangle\rangle$ and its time derivative (giving the instantaneous field radiation). Thus the radiative term $H_R(t)$ is merely the interaction of the molecular dipoles with their radiated field, and can be written as $H_R(t) = \mu^2 \vec{z} \cdot \frac{d\langle\langle\cos\theta\rangle\rangle(t)}{dt}$ with $\vec{z}$ the unit vector in the THz polarization direction (and the direction of orientation z).

Figure 4 shows the simulation results of molecular orientation with the typically used Hamiltonian $\hat{H} = \hat{L}^2/2I + \hat{H}_{int}$ (red) and the modified $\hat{H} = \hat{L}^2/2I + \hat{H}_{int} + \hat{H}_R$ (blue). For clarity of graphical presentation we chose a linear molecule with B=0.7cm$^{-1}$, D=3.5·10$^{-6}$cm$^{-1}$ at the temperature of 150K excited by a single-cycle THz field with FWHM=1.5ps and center frequency ω$_0$=0.6THz. We assume collision-free rotational evolution, thus no decay phenomena other than the coherent radiative emission are considered.

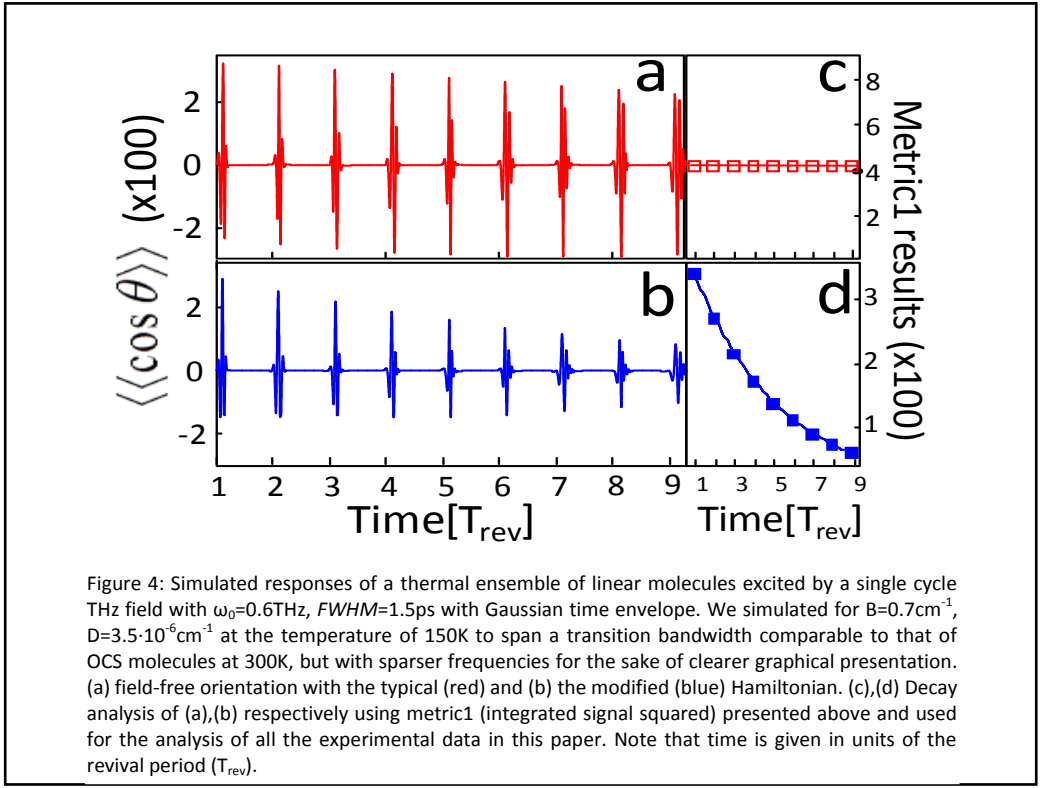

Figure 4: Simulated responses of a thermal ensemble of linear molecules excited by a single cycle THz field with ω$_0$=0.6THz, *FWHM*=1.5ps with Gaussian time envelope. We simulated for B=0.7cm$^{-1}$, D=3.5·10$^{-6}$cm$^{-1}$ at the temperature of 150K to span a transition bandwidth comparable to that of OCS molecules at 300K, but with sparser frequencies for the sake of clearer graphical presentation. (a) field-free orientation with the typical (red) and (b) the modified (blue) Hamiltonian. (c),(d) Decay analysis of (a),(b) respectively using metric1 (integrated signal squared) presented above and used for the analysis of all the experimental data in this paper. Note that time is given in units of the revival period (T$_{rev}$).

The radiative term, $\hat{H}_R$, included in Fig4.b results in an exponential decay of the degree of orientation, analyzed in Fig.4d via metric1. The typically used Hamiltonian retains the amplitude of the signals with no observed decay as expected (Fig4a and analysis in Fig.4c). Based on the suggested model, we expect the experimental radiative decay to follow a power dependence on the dipole moment, $\gamma_{Radiation} \propto \mu^a$. In the (theoretical) limit of pure collision-free dynamics (without incoherent energy dissipation as in collisions) and broad enough THz excitation that resonantly excites all of the rotational transitions, we expect a quadratic dependence $\gamma_{Radiation} \propto \mu^2$, namely *a=2*. Under our experimental conditions (pressure of few Torr and ambient temperatures) collisions play a significant role in the rotational decay and decoherence (manifested by $\gamma_{Align} > 0$ in Fig.3), and affect the observed dynamics in two ways: on one hand, collisions result in decoherence that decreases the signal amplitudes with time (as in optical-induced alignment), and on the other hand, decoherence reduces the coherent radiative emission. Therefore at sufficiently high pressures, the decay of both the alignment and orientation signals will be governed solely by collisions, with minimal or zero contribution of the radiative decay (at sufficiently high pressure, complete decoherence may be reached before the first event of emission, corresponding to a=0).

Note that while collisions occur continuously throughout the rotational evolution, the radiation emission occurs only at the specific times of orientation, i.e. at integer revival periods. Thus, in order to compare the radiative decay of different molecular species, one must consider $\gamma_{Radiation}$ in units of the rotational revival time (see the right-most column in Table.1). Although we have only three $\gamma_{Radiation}$ values (for the three gasses measured), we could not resist fitting the results to a power law $\propto \mu^a$, that yielded the value of a=1.75.

The presented theoretical model captures the basic features of the coherent radiative decay phenomenon discussed in this work. Clearly, the model should be further developed to include the effects of gas density and incoherent processes like collisions that affect the field-free evolution of the rotating ensemble. These will be studied in a forthcoming publication.

To conclude, we experimentally measured and compared between the decay of field-free THz-induced-orientation and optical-induced-alignment of three molecular species at a range of gas pressures. The difference between the decay rates of alignment and orientation is attributed to the loss of rotational energy via coherent emission of radiation. The experimental results are complemented by theory by including a radiative emission term in the typically used Hamiltonian. In this radiative term the electromagnetic emission is derived from the transient dipole oscillations of the ensembles, and act back on the molecules via the dipole interaction. While our work is focused on gas phase rotational dynamics and reveals the dramatic contribution of coherent radiative emission to the decay, the underlying phenomenon is general and should be considered and taken in account whenever field-free coherent dipole oscillations are induced.

Acknowledgment: The authors wish to thank Dr. Shimshon Kallush, Dr. Erez Boukobza and Prof. Uzi Even for stimulating discussions. This work is supported by the Israeli Science Foundation grant no.1065/14, the Marie Curie CIG grant no. 631628 and in part by the INREP grant (ISF) no.2797/11 and the Wolfson Foundation.

† Corresponding author: sharlyf@post.tau.ac.il

[1] H. Stapelfeldt and T. Seideman, *Rev. Mod. Phys.* **75**, 543 (2003).
[2] M. Lemeshko, R. V. Krems, J. M. Doyle, S. Kais, *Molecular Physics*, **111**, 1642 (2013).
[3] S. Fleischer, Y. Khodorkovsky, E. Gershnabel, Y. Prior, I. Sh. Averbukh, *Isr. J. Chem*. **52**, 414-437 (2012).
[4] I. Sh. Averbukh and N. F. Perelman, Phys. Lett. A 139 , 449 (1989).
[5] T. Seideman, *Phys. Rev. Lett.* **83**, 4971 (1999).
[6] S. De, I. Znakovskaya, D. Ray, F. Anis, N. G. Johnson, I. A. Bocharova, M. Magrakvelidze, B. D. Esry, C. L. Cocke, I. V. Litvinyuk, and M. F. Kling, *Phys. Rev. Lett.* **103**, 153002 (2009).
[7] E. Frumker, C. T. Hebeisen, N. Kajumba, J. B. Bertrand, H. J. Wörner, M. Spanner, D. M. Villeneuve, A. Naumov, and P. B. Corkum, *Phys. Rev. Lett.* **109**, 113901 (2012)
[8] O. Ghafur, A. Rouzée, A. Gijsbertsen, W. K. Siu, S. Stolte, and M. J. J. Vrakking, *Nature Phys.* **5**, 289 (2009).
[9] L. Holmegaard, J. H. Nielsen, I. Nevo, H. Stapelfeldt, F. Filsinger, J. Küpper, and G. Meijer, *Phys. Rev. Lett.* **102**, 023001 (2009).
[10]  A. Goban, S. Minemoto, and H. Sakai, *Phys. Rev. Lett.* **101**, 013001 (2008).
[11] S. Fleischer, Y. Zhou, R. W. Field, and K. A. Nelson, *Phys. Rev. Lett,* **107**, 163603 (2011).
[12] H. Harde, S. Keiding, and D. Grischkowsky, *Phys. Rev. Lett* **66**,1834 (1991).
[13] M. Machholm and N. E. Henriksen, *Phys. Rev. Lett.* **87**, 193001 (2001);

[14] J. Ortigoso, J. Chem. Phys 137, 044303 (2012).
[15] S. Fleischer, R.W. Field, and K. A. Nelson, arXiv:1405.7025v4 (2015).
[16] R. Damari, S. Kallush, and S. Fleischer, *Phys. Rev. Lett.* **117**, 103001 (2016).
[17] See supplementary information section [link]
[18] E. Hamilton, T. Seideman, T. Ejdrup, M. Poulsen, C. Z. Bisgaard, S. S. Viftrup, and H. Stapelfeldt, Phys. Rev. A 72, 043402 (2005)
[19] P. Babilotte, K. Hamraoui, F. Billard, E. Hertz, B. Lavorel, O. Faucher, and D. Sugny, Phys. Rev. A, 94, 043403 (2016).
[20] J. Lu, Y. Zhang, H. Y. Hwang, B. K. Ofori-Okai, S. Fleischer, and K. A. Nelson, *PNAS* **113**, 11800 (2016).
[21] N. Owschimikow, F. Konigsmann, J. Maurer, P. Giese, A. Ott, B. Schmidt, and N. Schwentner, *J. Chem. Phys.* **133**, 044311 (2010).
[22] I. F. Tenney, M. Artamonov, T. Seideman, and P. H. Bucksbaum, *Phys. Rev. A* **93**, 013421 (2016).
[23] S. Ramakrishna, and T. Seideman, *J. Chem. Phys.* **124**, 034101 (2006).
[24] S. Ramakrishna, and T. Seideman, *Phys. Rev. Lett.* **95**, 113001 (2005).
[25] Y. Khodorkovsky, U. Steinitz, J.M. Hartmann, and I. Sh. Averbukh, *Nature comm.* **6**, 7791 (2014).
[26] A. A. Milner, A. Korobenko, J. W. Hepburn, and V. Milner, *Phys. Rev. Lett.* **113**, 043005 (2014).
[27] B. Friedrich and D. Herschbach, *J. Phys. Chem.* **99,** 15686 (1995).
[28] M. Lapert and D. Sugny, *Phys. Rev. A* **85**, 063418 (2012).
[29] M. Machholm and N. E. Henriksen, *Phys. Rev. Lett.* **87**, 193001 (2001).
[30] N. E. Henriksen, *Chem. Phys. Lett*. **312**, 196-202 (1999).
[31] E. Gershnabel, I. Sh. Averbukh, and Robert J. Gordon, *Phys. Rev. A*, **73** 061401 (2006).
[32] K. Kitano**,** Nobuhisa Ishii, and Jiro Itatani**,** *Phys. Rev. A,* **84**, 053408 (2011).